\documentclass[aps,prl,nofootinbib,twocolumn,superscriptaddress]{revtex4}

\pdfoutput=1

\usepackage{hyperref}
\usepackage{epsfig}
\usepackage{setspace}

\usepackage{amsmath}
\usepackage{amssymb}
\usepackage{graphicx}
\usepackage{slashed}
\usepackage{multirow}

\usepackage{subfigure}

\newcommand{\ie}{{\it i.e.}}

\newcommand{\be}{\begin{equation}}
\newcommand{\ee}{\end{equation}}
\newcommand{\br}{\begin{eqnarray}}
\newcommand{\bea}{\begin{eqnarray}}
\newcommand{\eea}{\end{eqnarray}}
\newcommand{\er}{\end{eqnarray}}
\newcommand{\ba}{\begin{array}}
\newcommand{\ea}{\end{array}}
\newcommand{\bi}{\begin{itemize}}
\newcommand{\ei}{\end{itemize}}
\newcommand{\bn}{\begin{enumerate}}
\newcommand{\en}{\end{enumerate}}
\newcommand{\bc}{\begin{center}}
\newcommand{\ec}{\end{center}}

\newcommand{\Eq}[1]{Eq.~(\ref{#1})}
\newcommand{\rfn}[1]{(\ref{#1})}

\newcommand{\beq}{\begin{equation}}
\newcommand{\eeq}{\end{equation}}

\newcommand{\U}{\scriptscriptstyle U}
\newcommand{\D}{\scriptscriptstyle D}

\newcommand{\gsim}{\lower.7ex\hbox{$\;\stackrel{\textstyle>}{\sim}\;$}}
\newcommand{\lsim}{\lower.7ex\hbox{$\;\stackrel{\textstyle<}{\sim}\;$}}

\newcommand{\bs}{\begin{small}}
\newcommand{\es}{\end{small}}

\newcommand{\qui}{q_{{\scriptscriptstyle U}_{\!i}}}
\newcommand{\qdi}{q_{{\scriptscriptstyle D}_{\!i}}}

\newcommand{\eu}{e_{{\scriptscriptstyle U}}}
\newcommand{\ed}{e_{{\scriptscriptstyle D}}}

\newcommand{\BR}{{\rm BR}}
\def\mysection#1{{\bf #1.} }

\begin{document}

\title{
Dark photons and  resonant monophoton signatures in Higgs boson decays at the LHC}

\author{Emidio Gabrielli}
\affiliation{NICPB, Ravala 10, 10143 Tallinn, Estonia}
\affiliation{Dipart. di Fisica Teorica, Universit\`a di Trieste, Strada Costiera 11, I-34151 Trieste,  and INFN, Sezione di Trieste, Via Valerio 2, I-34127 Trieste, Italy}
\author{Matti Heikinheimo}
\affiliation{NICPB, Ravala 10, 10143 Tallinn, Estonia}
\author{Barbara Mele}
\affiliation{INFN, Sezione di Roma,  P. le A. Moro 2, I-00185 Rome, Italy}
\author{Martti Raidal}
%\email[]{martti.raidal@cern.ch}
\affiliation{NICPB, Ravala 10, 10143 Tallinn, Estonia}
\affiliation{Institute of Physics, University of Tartu, Ravila 14c, 50411 Tartu, Estonia}

\date{\today}
\begin{abstract}
Motivated by dark-photon $\bar{\gamma}$ scenarios extensively considered in the literature, we explore experimentally allowed models where the Higgs boson coupling to photon and dark photon $H\gamma\bar{\gamma}$ can be enhanced. 
 Correspondingly, 
large rates for the  $H\to \gamma\bar \gamma$ decay become plausible, giving rise to   one monochromatic  photon 
with $E^{\gamma}\simeq m_H/2$ (i.e., more than twice the photon energy  in  
the rare standard-model decay $H\to \gamma Z\to\gamma\bar\nu\nu$),
and a similar amount of missing energy. We  perform a model-independent study of this exotic resonant monophoton signature at 
 the LHC, featuring a distinctive $E^{\gamma}_T$ peak around 60 GeV,  and  $\gamma+\slashed{E}_T$ transverse invariant mass ruled by $m_H$. 
 At parton level, we find a $5\,\sigma$ sensitivity of the  present  LHC data set for a $H\to \gamma\bar{\gamma}$  branching fraction of $0.5\%$.
Such large branching fractions can be naturally obtained in dark $U(1)_F$ models explaining the origin and hierarchy of the standard model Yukawa couplings.
We urge the LHC experiments to search for this new exotic resonance in the present 
data set, and in future LHC runs.
\end{abstract}

%\pacs{}
%\keywords{}

\maketitle

\mysection{Introduction}
Although dark matter (DM) is five times more abundant in the Universe than ordinary baryonic matter~\cite{Ade:2013zuv}, its properties are yet unknown.
It is plausible that the dark sector, which is weakly coupled to the standard model (SM), possesses rich internal structure and interactions.
Among the most popular scenarios is the idea that the dark sector contains light or massless gauge bosons~\cite{Essig:2013lka}
that mediate long-range forces between dark particles. In cosmology the dark photons
may solve the small-scale structure formation problems~\cite{Spergel:1999mh,Aarssen:2012fx} and, for massless dark photons~\cite{Ackerman:mha}, 
predict  dark discs of galaxies~\cite{Fan:2013tia}.  In astroparticle physics dark photons may induce Sommerfeld enhancement of DM annihilation 
cross section needed  to explain the PAMELA-Fermi-AMS2 positron anomaly~\cite{ArkaniHamed:2008qn},  may
assist light DM annihilations to reach the phenomenologically required magnitude, 
and make asymmetric DM scenarios phenomenologically viable~\cite{Zurek:2013wia}. Dark/hidden photon scenarios have also been extensively considered in  
beyond-the-SM frameworks in particle physics \cite{Holdom:1985ag,Babu:1997st,Feldman:2007wj,
Abel:2008ai,Pospelov:2008zw,Chun:2010ve,Choi:2013qra}.

Recently, a new paradigm has been proposed for generating exponentially spread SM Yukawa couplings from unbroken $U(1)_F$ quantum numbers 
in the dark sector~\cite{Gabrielli:2013jka}. 
In this approach, nonperturbative  flavor- and chiral-symmetry breaking is transferred from the dark to visible sector via heavy scalar messenger 
fields~\cite{Gabrielli:2013jka,Ma:2014rua} that might give distinctive new physics (NP) signals at the LHC.
For massless dark photons~\cite{Ackerman:mha} the $U(1)_F$ kinetic mixing with $U(1)_Y$ can be tuned away~\cite{Holdom:1985ag} on shell,
in agreement with all existing constraints~\cite{Essig:2013lka}, while
off-shell contributions give rise to higher-dimensional contact operators 
strongly suppressed by the scale of the heavy messengers' mass.
 Therefore, in this scenario direct tests of dark photons 
 may require new ideas. 
 On the other hand, the photon kinetic mixing can induce millicharge couplings of dark fermions with ordinary photons, that can  already be probed at the LHC
\cite{Jaeckel:2012yz}. This could allow one to constrain some regions of the model parameter space.

In this work we show that, in the unbroken dark $U(1)$ scenarios, the Higgs-boson two-body 
decay $H\to \gamma\bar\gamma$ to one photon $\gamma$ and one dark photon $\bar\gamma$  can be enhanced despite  existing constraints, 
providing  a very distinctive NP signature of a single photon plus missing energy at the Higgs resonance. 
 If this signature will be discovered at the LHC, CP invariance will imply the spin-1 nature of the missing energy, 
 excluding  axions or other ultralight scalar particles. 

Monophoton plus $\slashed{E}_T$ signatures have been used by the LHC experiments to search for NP scenarios  such as
extra dimensions, supersymmetry, DM pair  production~\cite{Chatrchyan:2012tea}, and 
SM continuous $Z \gamma$ production~\cite{Aad:2013izg}. 
In those cases the photon and $\slashed{E}_T$ distributions are mostly monotonic and not much structured, corresponding to the 
nonresonant production of different invisible particles that carry away broadly distributed missing energy. 
A resonant monophoton plus $\slashed{E}_T$ signature occurs 
in the SM rare Higgs decays $H\to Z\gamma\to\bar\nu\nu\gamma$ with a $\gamma$ energy of about 30 GeV, which is much lower that the  
$m_H/2$ photon energy in $H\to \gamma\bar\gamma$. 
To our knowledge this exotic Higgs decay channel, giving rise to a striking experimental signature, has not been considered so far
(for a review of exotic Higgs signatures see~\cite{Curtin:2013fra,Falkowski:2014ffa}).
 The aim of this work is to show that the corresponding 
$\gamma\bar\gamma$ resonance can be realistically detected at the LHC, providing a nontrivial test of
dark-photon scenarios at the LHC.

Inspired by the model in~\cite{Gabrielli:2013jka}, we present a more general model-independent framework that can predict enhanced  
$H\to\gamma\bar\gamma$ decay rates.  We perform a parton-level Monte Carlo study of this process
versus relevant SM backgrounds, and show that, for a significant part of the model parameter space, this process could be observed at the LHC.
Detailed detector-level studies of the proposed signature will be needed to find the actual LHC sensitivity to massless dark-photon scenarios.

\mysection{Theoretical framework}
The aim of the model in \cite{Gabrielli:2013jka} is to explain the observed hierarchies in fermion masses, \ie, in the SM Yukawa couplings,
by exponential hierarchies due to quantum numbers of an exact new $U(1)_F$ gauge symmetry in the dark sector.
In this model  the hidden sector consists of dark fermions charged under  $U(1)_F$. 
As previously noted in \cite{Gabrielli:2007cp}, spontaneous chiral symmetry breaking can be triggered by the presence of a higher derivative kinetic term in the gauge sector, suppressed by a scale $\Lambda$, which can be interpreted as the mass scale of the associated Lee-Wick ghost~\cite{Lee:1971ix} of the $U(1)_F$ gauge theory.
The dark fermion masses $M_i$  can be dynamically generated via nonperturbative mechanism {\it a la}  Nambu-Jona-Lasinio~\cite{Nambu:1961tp}
 as a nontrivial solution of the (finite) mass-gap equation. 

The SM Yukawa couplings $Y_i$ are dynamically generated at one loop by the messenger fields that carry the SM quantum numbers of  squarks and sleptons of supersymmetric models.
In the approximation of a universal average mass $\bar{m}$ for the  messenger fields, we get
\bea
Y_i&=&Y_0(M_i/\bar{m})
\exp{\left(-\frac{2\pi}{3\bar \alpha q_i^2}\right)},
\label{Y}
\eea
where $\bar \alpha$ is the $U(1)_F$ fine structure constant and $q_i$ are dark fermion $U(1)_F$ charges. The loop function 
$Y_0(M_i/\bar{m})$ (see \cite{Gabrielli:2013jka}) 
has a weak dependence on $M_i/\bar{m}$, and  is proportional to 
$Y_0 \sim \langle S\rangle \Lambda/\bar{m}^2$,
where $\langle S\rangle$ is the vacuum expectation value (vev) 
of the singlet scalar field $S$ required to break the 
$H\to -H$ parity.
Equation \Eq{Y} implies that the origin of flavor in the SM Yukawa couplings resides in the  nonuniversality of the $U(1)_F$ charges in the dark sector.
Vacuum stability bounds and Eq.(1) require the average mass of colored messengers to be above 50 TeV~\cite{Gabrielli:2013jka}.
The dark fermions are the lightest dark particles which, due to $U(1)_F$, are 
all stable and can potentially contribute to the dark matter density of the universe. Because of the long-range $U(1)_F$ interaction with nonuniversal charges, the cosmology of the dark sector is nontrivial, and constraints apply on the masses and couplings of the dark fermions~\cite{Ackerman:mha}. We will not discuss the DM phenomenology further in this work.

An analogous model has been recently proposed in \cite{Ma:2014rua}, 
although  there the dynamics responsible for generating the 
hierarchy in the dark fermion spectrum  is missing.
We  then compute BR$(H\to\gamma\bar\gamma)$ in a model-independent way to extend  our results to all models of this type.

%\vskip 0.3cm
\mysection{Higgs decays to $H\to \gamma\, \bar{\gamma}$}
Consider a generic messenger sector like in \cite{Gabrielli:2013jka,Ma:2014rua}, consisting of
left doublet and right singlet scalars  $S_L^i$, $S_R^i$, with a flavor universal
mass term. The latter carry squark and slepton quantum numbers under the SM
gauge group, and additional $U(1)_F$ charges to couple to dark fermions.  Their couplings to the Higgs boson are (omitting the flavor indices)
\bea
{\cal L}^I_{MS} &=&
\lambda_S S \left(\tilde{H}^{\dag} S^{\U}_L S^{\U}_R+ H^{\dag} S^{\D}_L S^{\D}_R\right)
+ h.c. .
\label{LagMS}
\eea 
After the singlet $S$ scalar gets a vev,   a  $H\to \gamma\, \bar{\gamma}$ decay rate 
proportional to $\mu_S=\lambda_S \langle S\rangle$ is induced at one loop. 
After EWSB, the Lagrangian for generic $S_{L,R}$  is
\bea
{\cal L}^0_{S}&=& \partial_{\mu} \hat{S}^{\dag} \partial^{\mu}\hat{S} - \hat{S}^{\dag} M^2_S \hat{S},
\eea
where  $\hat{S}=(S_L,S_R),$ and the  mass term is given by
\begin{equation}
M^2_S = \left (
\begin{array}{cc}
m^2_L & \Delta \\
\Delta & m^2_R 
\end{array}
\right),
\label{M2}
\end{equation}
where $\Delta=\mu_S v$ parametrises the left-right mixing of scalars, and  
$v$ the SM Higgs vev.
Then, if  $\varepsilon_1^{\mu}(k_1)$ and $\varepsilon_2^{\mu}(k_2)$ are 
the photon and dark-photon polarization vectors, respectively,
we express the $H\to \gamma\, \bar{\gamma}$ amplitude   as 
\bea
M_{\gamma\bar{\gamma}}&=& \frac{1}{\Lambda_{\gamma\bar{\gamma}} }\, T_{\mu\nu}(k_1,k_2) 
\varepsilon_1^{\mu}(k_1) \varepsilon_2^{\nu}(k_2),
\label{Mgg}
\eea
where $\Lambda_{\gamma\bar{\gamma}}$ parametrizes the effective scale associated
to the NP, and 
$T^{\mu\nu}(k_1,k_2)\equiv g^{\mu\nu} k_1\cdot k_2 -k_2^{\mu}k_1^{\nu}\, $. The total width is then
\bea
\Gamma(H\to \gamma \bar{\gamma})=m_H^3/(32\, \pi\, 
\Lambda_{\gamma\bar{\gamma}}^2).
\eea
If we neglect the Higgs boson mass  with 
respect to the messenger masses $m_{L,R}$ in the loop,  we obtain 
\bea
\frac{1}{\Lambda_{\gamma\bar{\gamma}}}\! \!&=&\! 
\frac{\mu_S \sqrt{\alpha \bar{\alpha}} R}{12 \pi} \!±\left(\!\frac{
\sqrt{(m_L^2-m_R^2)^2+4\Delta^2}}{m_L^2m_R^2-\Delta^2}\right)\!\sin{2\theta},
\eea
where $R=N_c \sum_{i=1}^3\left(\eu\qui+\ed\qdi\right)$, 
with $\qui, \qdi$ the $U(1)_F$ charges in the up and down sectors,
and $\eu=\frac{2}{3}$, $\ed=-\frac{1}{3}$ the corresponding EM charges;  
$\alpha$ is the EM  fine structure constant,
$N_c=3$ is the number of colors, and $\theta$ is the mixing angle diagonalizing Eq.~\rfn{M2}. 
The above result can be easily generalized to 
include the contributions of messengers in the leptonic sector, in this case $N_c=1$, $e_U=0$ and $e_D=-1$.
We  assume mass universality for 
$S_L$ and $S_R$, $m_L\simeq m_R\equiv \bar{m}$, giving $\theta=\pi/4$.
Defining $\xi = \Delta/\bar{m}^2$, the eigenvalues of \Eq{M2} become
%\bea
$
m^2_{\pm}=\bar{m}^2\left(1\pm \xi\right),
%\eea
$
 and the $\Lambda_{\gamma\bar{\gamma}}$  scale simplifies to 
\bea
\Lambda_{\gamma\bar{\gamma}} &=& \frac{6\pi v}{R\sqrt{\alpha\bar{\alpha}}}
\frac{1-\xi^2}{\xi^2}\, .
\eea
To avoid tachyons, one needs  $0\le \xi \le 1$.

The messengers induce new contributions also to the Higgs decays $H\to \gamma \gamma$ and $H\to \bar{\gamma} \bar{\gamma}$. The 
corresponding amplitudes have the same structure as  \rfn{Mgg}, and we obtain 
\bea
\Lambda_{\gamma\gamma} = \Lambda_{\gamma\bar{\gamma}}\, \frac{R}{R_0}
\sqrt{\frac{\bar{\alpha}}{\alpha}}\, , ~~~~
\Lambda_{\bar{\gamma}\bar{\gamma}} = \Lambda_{\gamma\bar{\gamma}}
\sqrt{\frac{\alpha}{\bar{\alpha}}}\frac{R}{R_1}\, ,
\eea
where $R_0=3 N_c(\eu^2+\ed^2)$, and $R_1=N_c \sum_{i=1}^3\left(\qui^2+\qdi^2
\right)$.

A   model-independent parametrization for the branching ratios (BRs) of the decays $H\to \gamma\, \gamma$, $H\to \gamma\, \bar{\gamma}$, and $H\to \bar{\gamma}\, \bar{\gamma}$ can be expressed as follows
\bea
\BR_{\gamma\gamma}\!\!&=&\!\!N \left(1\pm \sqrt{r_{\gamma\gamma}}\right)^2 \!,\;\;\;
\BR_{AB}=\,N r_{AB} ,
\label{BRS}
\eea
where $AB\equiv \{\gamma\bar{\gamma},\, \bar{\gamma}\bar{\gamma}\}$,
$N=\BR^{\rm SM}_{\gamma\gamma}/(1+r_{\bar{\gamma}\bar{\gamma}}\BR^{\rm SM}_{\gamma\gamma})$, and the ratios $r_{AB}$ are given by
\bea
r_{\gamma\bar{\gamma}}=2\,r_{\gamma\gamma} \frac{R^2}{R_0^2}
\left(\frac{\bar{\alpha}}{\alpha}\right)\, ,~~~
r_{\bar{\gamma}\bar{\gamma}}=r_{\gamma\gamma} \frac{ R_1^2}{R_0^2}
\left(\frac{\bar{\alpha}}{\alpha}\right)^2 ,
\eea
%with $r_{\gamma\gamma}$ defined as
where 
%\bea
$r_{\gamma\gamma}\equiv \Gamma^{\rm NP}_{\gamma\gamma}/
\Gamma^{\rm SM}_{\gamma\gamma}$.
Here $\Gamma_{\gamma\gamma}^{\rm NP}$ and $\Gamma_{\gamma\gamma}^{\rm SM}$ 
corresponds to the $H\to \gamma\gamma$  decay widths, 
mediated by new particles and SM ones, respectively. The $\pm$ 
signs  in Eq.(\ref{BRS}) 
corresponds to the constructive or destructive interference with the SM amplitude. 
In the scenario \cite{Gabrielli:2013jka}, the sign in $\BR_{\gamma\gamma}$
is predicted to be positive, while the corresponding value 
for $r_{\gamma\gamma}$ is given by
\bea
r_{\gamma\gamma}=\left(\frac{R_0\xi^2}{3 F\left(1-\xi^2\right)}\right)^2\, ,
\eea
where $F$ is the SM contribution, 
given by $F=F_W(\beta_W)+\sum_f N_c Q^2_f F_f(\beta_f)$,
with $\beta_W=4M_W^2/m_H^2$, $\beta_f=4m_f^2/m_H^2$, and $F_W(x)$ and $F_f(x)$ can be found in \cite{Marciano:2011gm}. 
Once the corresponding Higgs BRs are measured, the  $U(1)_F$ charges $q_i$ can be derived from the Yukawa couplings by Eq.~(\ref{Y}).
%=================================

\begin{figure}[t]
\begin{center}
\vskip -2.6cm
\hskip -2cm \includegraphics[width=0.59\textwidth]{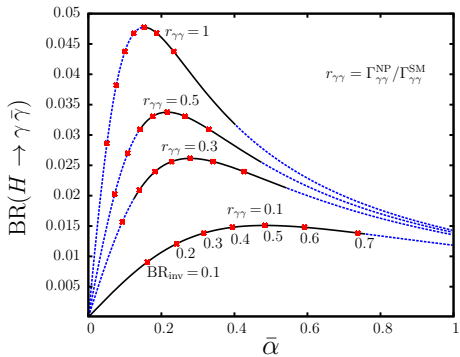} 
 \vskip -6.8cm
\caption{Predictions for $\BR(H\to \gamma \bar{\gamma})$ as functions of
 $\bar \alpha$ for different $\BR_{\rm inv}$ and $r_{\gamma\gamma}$  in the minimal model.
}
\label{BR}
\end{center}
\vskip -0.3cm
\end{figure}

To quantify predictions of this scenario,  in  Fig.~\ref{BR}  we plot 
$\BR(H\to \gamma \bar{\gamma})$ as a function of $\bar \alpha$, assuming that there is only one messenger contributing,   
with a charge $e=q=1$. The curves are evaluated for 
$r_{\gamma\gamma}=0.1,\, 0.2\, , 0.5\, , 1$.
The red dot bullets correspond to  different $\BR_{\bar{\gamma}\bar{\gamma}}$ values 
(or  Higgs invisible branching ratios BR$_{\rm inv}$), as shown in the plot (in the experimentally allowed range~\cite{CMS:2013bfa}).
%%%%comment
%%% we are assuming no other contribution to BR_inv
The full lines correspond to the interval 
$ \BR^{\rm SM}_{\gamma\gamma}/2\le \BR_{\gamma\gamma}\le 2 \;
\BR^{\rm SM}_{\gamma\gamma},$
where $\BR^{\rm SM}_{\gamma\gamma}=2.28\times 10^{-3}$, while the dashed lines correspond to predictions outside that range.
 We find that the signal $\BR(H\to \gamma \bar{\gamma})$  can be as large as 5\% (that is more than one order of magnitude larger than  $\BR^{\rm SM}_{\gamma\gamma}$), consistently with all model parameters and the LHC constraints.

We stress that large values of the messenger mixing-mass parameter 
$\xi$ are natural in the present scenario, in order to 
generate a large top-quark Yukawa coupling radiatively, and all EW precision tests can be satisfied due to the heavy and flavor universal
messenger sector \cite{Gabrielli:2013jka}.
In addition, large values of $\bar{\alpha}\gg \alpha$ are naturally expected 
in this scenario from Eq.(\ref{Y}), provided the splitting among the $q_i$ charges is not too small.
Consequently, the relatively large  $\BR(H\to \gamma \bar{\gamma})$ shown in Fig.~\ref{BR} can be considered a generic prediction of the present theoretical framework.\footnote{Large values of the mixing parameter $\xi$ can be safely
generated from the purely EW messenger sector, since the latter does not affect the Higgs production cross section in gluon fusion.}

%\vspace{0.1cm}
\mysection{Model independent analysis of $H\to \gamma \bar{\gamma}$ at the LHC}
The process $pp\to H \to \gamma\bar{\gamma}$ gives rise to the signal $\gamma + \slashed{E}_T$, where  $E_{\gamma}=m_H/2$ in the Higgs rest frame. In the lab frame,  one can define the  variable $M_T$, that is the transverse invariant mass of the $\gamma + \slashed{E}_T$ system, as
\begin{equation}
M_T=\sqrt{2p_T^\gamma \slashed{E}_T(1-\cos\Delta\phi)},
\end{equation}
where $p_T^\gamma$ is the photon transverse momentum, and $\Delta\phi$ is the azimuthal distance between the photon momentum and the missing transverse momentum $\slashed{E}_T$.

\begin{center}
\begin{figure}
\includegraphics[width=0.4\textwidth]{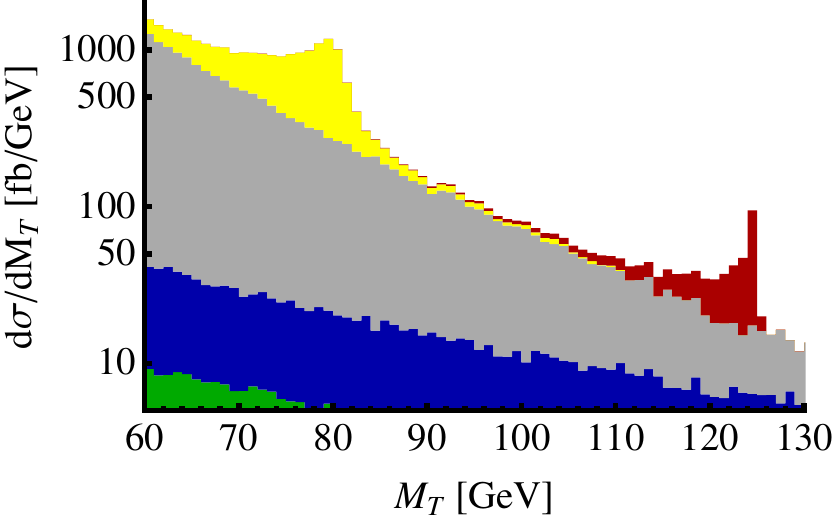} \vskip -0.3cm
\caption{The  $\gamma+\slashed{E}_T$ transverse invariant mass
 distribution (in fb/GeV) of the signal (red), and the main backgrounds $\gamma j$ (grey), $\gamma Z$ (blue), $j Z$ (green), and $W$ (yellow). 
For illustration, we show the signal for $BR(H\to \gamma\bar{\gamma})=5\%$.}
\label{MT_PT_distributions}
\end{figure}
\end{center}
\vspace{-1cm}

Like in the $W\to e\nu$ production, the $M_T$ observable features a narrow peak at the mass of the original massive particle (that is  $m_H$, see Fig.~\ref{MT_PT_distributions}).
 Also the $p_T^\gamma$ distribution will exhibit a similar structure around $m_H/2$. These features  allow for a very efficient cut-based search strategy,  looking for events with a single photon and missing energy, with no jets or leptons, and cutting around the expected maximum of the $M_T$ and  $p_T^\gamma$ distributions. These peaks could be relatively easy to pinpoint on top of the continuous relevant  backgrounds, for sufficiently large $H\to\gamma\bar\gamma$ decay rates.
Thus we formulate the criteria for event selection as follows:
\begin{itemize}
\item One isolated photon with $50\ {\rm GeV} < p_T^\gamma < 63\ {\rm GeV}$ and $|\eta^\gamma|<1.44$.
\item Missing transverse momentum with $\slashed{E}_T > 50$ GeV.
\item Transverse mass in  $100\ {\rm GeV} < M_T < 126\ {\rm GeV}$.
\item No isolated jets or leptons.
\end{itemize}
The most relevant backgrounds for the above selection criteria are, in order of importance:
\begin{enumerate}
\item $pp\to \gamma j$, 
%where the jet is missed and observed as missing energy;
 where large apparent $\slashed{E}_T$
 is created
by a combination of real $\slashed{E}_T$
 from neutrinos in
heavy quark decays and mismeasured jet energy.
\item $pp\to \gamma Z \to \gamma \nu\bar{\nu}$   (irreducible background);
\item $pp\to j Z \to j \nu\bar{\nu}$, where the jet is misidentified as a photon;
\item $pp\to W \to e\nu$, where the electron (positron) is misidentified as a photon;
\item $pp\to \gamma W \to \gamma \ell \nu$, where the lepton is missed;
\item $pp\to \gamma \gamma$, where one of the photons is missed.
\end{enumerate}

The $pp\to \gamma j$ background is expected to be dominant for the $\slashed{E}_T$ range
 relevant here, and also the most difficult to estimate without detailed information about the detector performance \cite{CMS:vgm}.  We have evaluated this background by simulating events with one photon and one jet,  treating jets with $|\eta|>4.0$ as missing energy, following \cite{Petersson:2012dp} (a  more detailed investigation of the $pp\to \gamma j$ background, although crucial for assessing the actual experiment potential, is beyond the scope of this work). All the other backgrounds have also been estimated through a 
 parton-level simulation, expected to be relatively accurate for  electroweak processes (applying a probability $10^{-3}$ and 1/200 to misidentify a jet and an electron, respectively,  as a photon). We will neglect the subdominant backgrounds from processes 5 and 6 (the $H\to \gamma\gamma$ background is also negligible). The contribution of relevant backgrounds passing the cuts is shown in Table~\ref{signal_vs_background}, and the scaling of the different components with the transverse mass is shown in Fig.~\ref{MT_PT_distributions}. 
Although our leading-order parton-level analysis, after applying a cut on  $p_T^\gamma$ is not much affected by a further cut on the $M_T$ variable, we expect the latter to be very effective in selecting  our structured signal over the continuous reducible QCD background \cite{CMS:vgm}.

\begin{table}[t]
\begin{center}
\begin{tabular}{c|c|c}
 & $\sigma\times A_1$ & $\sigma\times A_2$ \\ \hline
Signal $\BR_{H\to \gamma\bar{\gamma}}=1\%$ & 65 & 34 \\ \hline
 $\gamma j$ & 715 & 65 \\
 $\gamma Z\to \gamma \nu\bar\nu$ & 157 & 27 \\
 $j Z\to  j \nu\bar\nu$ & 63 & 11 \\
 $W\to e\nu$ & 22 & 0 \\ \hline
Total background & 957 & 103 \\ \hline
$S/\sqrt{S+B}$  ($\BR_{H\to \gamma\bar{\gamma}}=1\%$) & 9.1 & 13.0 \\
$S/\sqrt{S+B}$  ($\BR_{H\to \gamma\bar{\gamma}}=0.5\%$)& 4.6 & 6.9 \\
\end{tabular}
\end{center}
\vskip -0.5cm
\caption{The cross section times acceptance (in fb) for the signal and background processes at 8 TeV for the  selections  ($A_1$) $50\ {\rm GeV} < p_T^\gamma < 63\ {\rm GeV}$;  ($A_2$) $60\ {\rm GeV} < p_T^\gamma < 63\ {\rm GeV}$. In all cases $|\eta^\gamma|<1.44$,  and  $S/\sqrt{S+B}$ is for 20 fb$^{-1}$. The significance improves with tighter cuts, but this is subject to experimental resolution and radiative corrections.}
\label{signal_vs_background}
\end{table}
With the existing data set of $20\ {\rm fb}^{-1}$, for $\BR(H\to \gamma \bar{\gamma})=1\%$, we get a significance $S/\sqrt{S+B}$ of 9 standard deviations $(9\sigma)$,  
with $S (B)$ the number of signal (background) events passing the cuts. The sensitivity limit for a $5\,\sigma$ discovery is then estimated to be  $\BR(H\to \gamma \bar{\gamma})\sim 0.5\%$   with the existing dataset.

\vskip 0.3cm
\mysection{Conclusions}
Motivated by possible cosmological and particle physics hints for the existence of massless dark photon $\bar\gamma$, 
we have performed a model-independent study of the exotic $H\to\gamma\bar \gamma$  decay. 
At the LHC this results in a single photon plus $\slashed{E}_T$ signature, with both energies peaked at $m_H/2$.
At parton level,  we estimate that a $5\,\sigma$ discovery can be reached with the existing 8 TeV LHC data sets if  BR($H\to \gamma\bar{\gamma})\sim 0.5\%$. Such a large branching ratio can be easily obtained in dark $U(1)_F$ models explaining the origin and hierarchy of the SM Yukawa couplings.
The proposed experimental signature is new, and requires detailed detector-level studies  to draw realistic conclusions on the LHC sensitivity to dark photons.

\mysection{Acknowledgment}
We thank S.~Chauhan, J.P.~Chou and J. Alcaraz Maestre for communications, and C.~Spethmann for collaboration in the early stages of the project.  
This work was supported by  grants MTT60,  IUT23-6, CERN+, and by EU through the ERDF  CoE program.

\end{document}